# Forbidden triads and Creative Success in Jazz: The Miles Davis Factor[1]

Balazs Vedres
*Central European University*

This article argues for the importance of forbidden triads – open triads with high-weight edges – in predicting success in creative fields. Forbidden triads had been treated as a residual category beyond closed and open triads, yet I argue that these structures provide opportunities to combine socially evolved styles in new ways. Using data on the entire history of recorded jazz from 1896 to 2010, I show that observed collaborations have tolerated the openness of high weight triads more than expected, observed jazz sessions had more forbidden triads than expected, and the density of forbidden triads contributed to the success of recording sessions, measured by the number of record releases of session material. The article also shows that the sessions of Miles Davis had received an especially high boost from forbidden triads.

## Introduction

While the spread and adoption of innovations has been a central concern of social network research (Centola 2015), we know much less about the network structures that contribute to the generation of innovation. In the efforts to understand how new ideas are incorporated into existing practices, much attention was devoted to the duality of network closure and open weak ties. Closure is seen to provide a trusted collaborative environment to realize ideas captured by weak ties from outside the cohesive core (Burt 2005). This paper is about the creative potential of a third triad type: forbidden

[1] Forthcoming in *Applied Network Science*. Research for this article was supported by the National Science Foundation (award number 1123807). This article builds on concepts developed in prior publications with David Stark and Mathijs deVaan. I am thankful for their assistance and helpful comments, as well as for valuable insights of Albert-Laszlo Barabasi, Janos Kertesz, Levente Littvay, Thomas Rooney, Roberta Sinatra, Michael Szell, Orsolya Vasarhelyi, Johannes Wachs, and two anonymous reviewers. I am also thankful for the insights and assistance of Carl Nordlund in collecting, preparing, and cleaning the data.

triads – connected triplets with two strong ties and one open dyad. Such triads provide opportunities to recognize unique combinations of skills: generative combinations of tacit knowledge that evolved in the two intersecting strong-tie dyads.

While weak ties and closure are building blocks of small world structures, forbidden triads are building blocks of fold networks. Fold networks have been identified as predictors of the generation of novelty, where overlapping cohesive communities contribute to both the recognition of a novel possibility, and the realization of the novel idea as a product (De Vaan, Stark, and Vedres 2015). It was also demonstrated that it is not the agency at the overlap of communities that matters, but the successful mobilization of the non-intersecting unique part of overlapping communities, and overlapping communities have also been shown to be more unstable than non-overlapping communities (Vedres and Stark 2010). Fold networks operate by a generative tension: They provoke the generation of novelty, but also contribute to coordination and loyalty conflicts.

While the overlapping of communities provides a clear mechanism for realizable novelty, the empirical operationalization of fold networks has so far been cumbersome. One needs to identify communities first, and especially communities that overlap to measure network folding. Previous studies have used the number of community memberships of a given node (Vedres and Stark 2010), or the number of subgroup overlaps within a larger collective (De Vaan et al. 2015). But community detection is far from being a universal and simple process: The large number of community detection algorithms is a symptom of the complex nature of mapping the community concept (especially with overlaps allowed) onto sets of nodes in a network (Granell et al. 2015; Xie, Kelley, and Szymanski 2013).

This article uses the density of forbidden triads as a direct measure of foldedness. Forbidden triads are open triads with high-weight legs. The concept was introduced by Mark Granovetter in his seminal article on the significance of weak ties in predicting labor market success (Granovetter 1973). Granovetter argued that strong ties are most likely to be closed, and weak ties are the ones that can bridge communities, and thus provide access to diverse information. Granovetter also argued that strong and open ties are rare and fleeting, thus the label "forbidden."

Up to this point no one has analyzed the importance of forbidden triads for success, as the four-decades long tradition of parsing triads into strong-and-closed and weak-and-open left little room for seeing any significance for forbidden triads. Prior research has discussed the negative consequences of overlapping closed triads, under the label of Simmelian brokerage, or reinforced brokerage (Burt 2015; Krackhardt and Kilduff 2002) in the context of business ties and managerial networks.

I argue, however, that forbidden triads are the key building blocks of creative networks, as these triads are occasions where two high-weight edges intersect on a central node. This creates an opportunity to hybridize prior practices (such as musical styles), to generate a new kind of practice. Forbidden triads go together with overlapping cohesion: there are many forbidden triads around the intersecting nodes of two communities, if one accepts the assumption that triadic closure is related to high edge weight. In a network with folds (overlapping communities) one expects to see a high density of forbidden triads. I define triad density as the proportion of a kind of triad to all connected triads.

The only decision to be made in defining a forbidden triad is the threshold for an edge weight to qualify as a strong tie. In this article I use a minimal threshold of repeated (as opposed to one-shot) interaction, but also analyze various thresholds to estimate sensitivity of statistical models to classifying strong ties.

This article uses jazz as a case of a creative field, where collective innovation is intricately linked with artistic value (Phillips 2013). Jazz is a field where the product – the music – is a result of a team effort, and band leaders exert considerable effort to select and combine musicians (Kirschbaum and Ribeiro 2016). Miles Davis – arguably the most important band leader in the history of jazz – was especially vocal about the importance of combining musicians for fresh approaches and new sounds (Davis and Troupe 1989). A prior analysis has shown that disconnectedness at the level of cities contributes to the evolution of new styles (Phillips 2011). In this article I take a similar approach at the level of the session, to show the creative importance of the lack of closure.

Using data on the entire history of recorded jazz from 1896 to 2010 (175,000 jazz sessions) the paper presents evidence for the contribution of forbidden triads in musical collaboration to success measured in the number of album releases. Forbidden triads are conceptualized as the proportion of connected triads in a session, where both connected legs of the triad are strong, and one dyad is unconnected. The threshold for tie strength was set at two shared session plays in the past. I analyze the sessions of Miles Davis separately as well, to show how forbidden triads added to the success of his sessions.

I test several hypotheses derived from the argument that creative success is a function of the density of forbidden triads. First, I test the hypothesis that edge strength relates positively to probability of closure. More precisely, I test the hypothesis that minimal triplet legs weight is positively related to the probability of closure. Second, I test the hypothesis that in the observed jazz world edge weights contribute less to the increase in the probability of closure, compared to a randomly rewired jazz world. In other words, jazz musicians tolerate openness of higher weight triads more than expected. I constructed randomly rewired jazz worlds that follow the principle of objectively possible counterfactuals. I generated random jazz worlds that could have happened within hard constraints that one can read from the data, and from practices of jazz musicians. Third, I test the hypothesis that forbidden triads are over-represented in observed jazz sessions, compared to jazz sessions in the rewired jazz worlds. Finally, I test the hypothesis that forbidden triads contribute to success at the session level, even if we take all conceived alternative explanations into account, including unobserved heterogeneity related to the identity of the band leader.

**Measuring collaboration and triad densities in Jazz**

To cover collaborations in recorded jazz, I used the Tom Lord Discography (Lord 2010), what is considered to be the most comprehensive source (Charry 2005). This discography collects musician participation in recording sessions from the beginnings of jazz to the present. I collected data up to 2010 about the complete set of recording sessions available in this discography. The core data is a tripartite graph of time-stamped instances (with a yearly time resolution) of sessions, musicians, and instruments.[2] The resulting dataset contains information about 175 064 recording sessions, taking place between 1896 and 2010. There were a total of 42 929 band leaders (or band names) and 187 784 musicians playing in these sessions, playing 11 940 different instruments[3]. The weighted collaborative network among musicians for a given session was generated by summing the prior co-plays for each musician dyads. To ensure strict temporal ordering, only sessions from year $t$-1 backwards were included, if the focal session took place in year $t$.

Figure 1 shows a sample session record from the Tom Lord Discography. The top of the data entry shows the session's ID, and the band leader (Charlie Parker in this case). Then the set of musicians is shown with their instruments. (In the case of Miles Davis the abbreviation "tp-1" means that he played trumpet on the first track.) After the musicians the place and date of recording is shown, and tracks are listed. Releases were coded from the appearance of unique catalog numbers at specific tracks, or for the whole session.

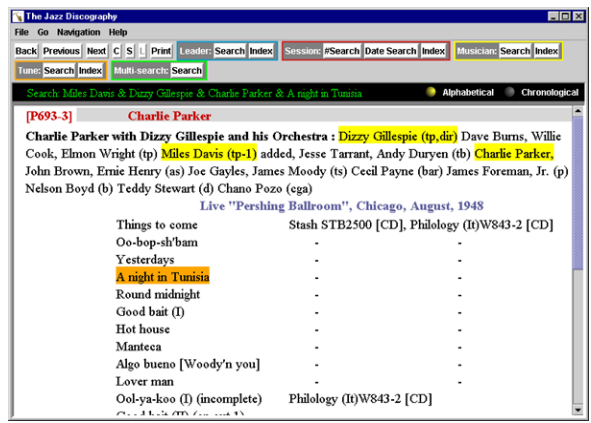

Figure 1: Sample of a session's record in the Tom Lord Jazz Discography. Source: https://lordisco.com/cdrom.html.

---

[2] The time resolution of sessions was set to the year of recording. Though most sessions had the recording day available, in several instances the day was missing, or was set to an obviously estimated date (like January 1st).
[3] The high number of different kinds of instruments indicate the innovative nature of the jazz field. As a comparison, the largest collection of musical instruments, the MIMO project (Musical Instrument Museums Online; mimo-international.com) lists a total of 1140 different kinds of instruments – an order of magnitude less. The current nomenclature of 11 940 instruments is a result of weeks of careful coding and collapsing. Jazz musicians have played everything from car parts to kitchen sinks and digital noises.

I am interested in relating the presence of three kinds of triads to levels of success. The first kind of triad is a building block of fold networks, the forbidden triad: a triad with two strong ties, and one absent tie. The second kind of triad is an open triad: two weak ties and one absent tie. The third kind of triad is closure: all three ties are present (of any strength). To measure the presence of these triad types I categorically delimit forbidden triads from open triads by a threshold tie strength value, and then count the number of three mutually exclusive triad types (forbidden triads, open triads, and closed triads). I then normalize the number of triads by the number of connected triads (with at least two ties present).

As an example consider Figure 2, that presents collaboration data from the "Kind of Blue" session with Miles Davis as band leader. The resulting album, "Kind of Blue" is the undisputed pinnacle of jazz: the most influential, most mentioned, and most re-issued jazz album in the history of the genre. This example shows the first of two sessions, from March 2, 1959. I used personnel lists from prior sessions to record the number of times pairs of musicians had played with each other. For example, considering panel a. of Figure 2, the dataset contains a total of 58 session plays for Paul Chambers prior to this session, 22 with Miles Davis (up to the end of 1958).

|   |   | 1 | 2 | 3 | 4 | 5 | 6 |
|---|---|---|---|---|---|---|---|
| 1 | Paul Chambers (bass) | (58) | 35 | 12 | 8 | 22 | 13 |
| 2 | John Coltrane (tenor sax) |  | (35) | 1 | 7 | 16 | 5 |
| 3 | Wynton Kelly (piano) |  |  | (25) | 11 | 0 | 1 |
| 4 | Jimmy Cobb (drums) |  |  |  | (24) | 6 | 10 |
| 5 | Miles Davis (trumpet) |  |  |  |  | (23) | 8 |
| 6 | Cannonball Adderley (alto sax) |  |  |  |  |  | (20) |

a: musician co-plays (diagonals are total plays)

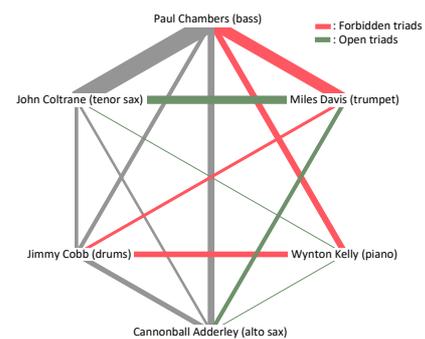

b: graph of co-plays

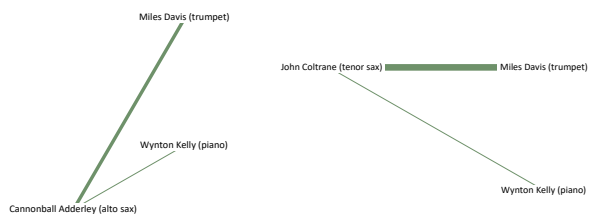

c: open triads

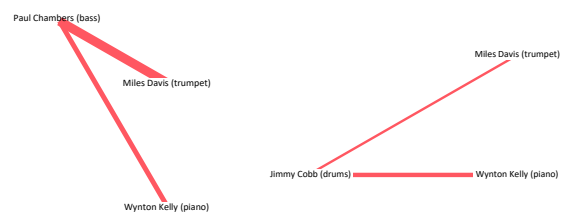

d: forbidden triads

Figure 2: Collaboration graph in the example session of "Kind of Blue", March 2, 1959.

The graph of collaborations on panel b. of Figure 2 shows open and forbidden triads, and one can see that these resulted from the missing link between Miles Davis and Wynton Kelly. Panels c. and d. show only the open and the forbidden triads separately.

**Triad types and minimal triplet legs weight**

To chart the space of connected triad types, I use two dimensions: minimal triplet legs weight and the probability of closure. Triplet legs are the two stronger edges in the connected triad, $w_{(3)}$ and $w_{(2)}$, the third and second edge weights in ascending order of triad edge weights. (The subscripted parenthesized numbers indicate the ascending order statistic.) In an open triad, the weakest edge $w_{(1)} = 0$. Minimal triplet legs weight is $w_{(2)}$, the second weight in ascending order, that captures the strength of the triad edges, independently of closure. This measure follows the same considerations that underlie the development of weighted clustering coefficients, where the geometric mean of edge weights is used, as it is more robust to outlying weight values (Onnela et al. 2005; Opsahl and Panzarasa 2009; Saramäki et al. 2007). I use the minimal weight instead of the geometric mean, because the goal is to distinguish triads where leg weights are higher than a threshold value. Weights here are raw ones, measuring the number of times two musicians recorded together in the studio.

Using the combination of minimal triplet leg weight and the presence of closure allows us to distinguish three triad types: Open triads are those where the minimal triplet leg weight is equal to one, and there is no closure. This triad is the building block of weak ties that connect communities (Granovetter 1973), or that contribute to lowering path length in small world networks (Watts 1999), and present brokerage opportunities (Burt 1992). Closed triads are those with all three edges present, regardless of the strength of tie. These are the triads that communities are built from where trust and shared values dominate (Burt 2005). Forbidden triads are those where the edge weights are higher than in the case of open triads, yet the triad is open. These triads are building blocks of fold networks, that are seen to promote creativity and creative success (Vedres and Stark 2010). Table 1 summarizes the classification scheme.

|  | Minimal legs weight | Closure |
|---|---|---|
| Open triad | $w_{(2)} = 1$ | $w_{(1)} = 0$ |
| Closed triad | $w_{(2)} > 0$ | $w_{(1)} > 0$ |
| Forbidden triad | $w_{(2)} > 1$ | $w_{(1)} = 0$ |

Table 1: Definition of triad types

Returning to the example of the "Kind of Blue" session, we see that it contains two forbidden triads (10% of triads), two weak triads (10%) and 16 closed triads (80%). Table 2 lists triads in these three categories, ranked by minimal legs weight. The interesting aspect of the "Kind of Blue" example, is that Wynton Kelly was invited to play the piano, even though there were no prior sessions where Miles Davis and Kelly played together. However, Kelly was no stranger to most of the musicians on the session – he had played 12 sessions with Chambers, and 11 with Cobb – both can be considered fairly strong ties. Davis met Kelly shortly before the "Kind of Blue" sessions, and brought him on board alongside the trusted pianist of the band, Bill Evans. Although Kelly played on only one tune on the album, "Freddy Freeloader", Miles commended the value of Kelly as an energizer in the band: "Wynton's the light for a cigarette. He lights the fire and he keeps it going[4]."

|  | *i* | *j* | *w<sub>ij</sub>* | *i* | *k* | *w<sub>ik</sub>* | *j* | *k* | *w<sub>jk</sub>* | *w<sub>(3)</sub>* | *w<sub>(2)</sub>* | min legs weight |
|---|---|---|---|---|---|---|---|---|---|---|---|---|
| Forbidden triads | 3 | 4 | 11 | 3 | 5 | 0 | 4 | 5 | 6 | 11 | 6 | 6 |
|  | 1 | 3 | 12 | 1 | 5 | 22 | 3 | 5 | 0 | 22 | 12 | 12 |
| Open triads | 3 | 5 | 0 | 3 | 6 | 1 | 5 | 6 | 8 | 8 | 1 | 1 |
|  | 2 | 3 | 1 | 2 | 5 | 16 | 3 | 5 | 0 | 16 | 1 | 1 |
| Closed triads | 2 | 3 | 1 | 2 | 6 | 5 | 3 | 6 | 1 | 5 | 1 | 5 |
|  | 2 | 4 | 7 | 2 | 6 | 5 | 4 | 6 | 10 | 10 | 7 | 7 |
|  |  |  |  |  |  | ... |  |  |  |  |  |  |
|  | 1 | 2 | 35 | 1 | 6 | 13 | 2 | 6 | 5 | 35 | 13 | 13 |
|  | 1 | 2 | 35 | 1 | 5 | 22 | 2 | 5 | 16 | 35 | 22 | 22 |

Table 2: Triad types in the example session of "Kind of Blue", March 2, 1959.

Further data tables at the session level record the identity of band leader, and the number of releases on which material recorded in the session appeared. The number of releases were coded as the number of unique catalog numbers that appear on the session's entry in the Tom Lord Discography.

**Rewired jazz worlds**

To test null hypotheses that relate only marginal frequencies to outcomes, I generated random jazz worlds by rewiring the observed tripartite dataset (of sessions, musicians, and instruments). The rewiring was accomplished following the principle of objective possibility: I was re-allocating musicians to sessions in a way that could have happened in real life (with, albeit, small likelihood), and I avoided composing sessions that were not possible according to available evidence. In other words, I was generating jazz worlds with sessions that the recording companies could have recorded.

The first principle for rewiring was that the number of musicians in the session needed to be preserved. Recording in the studio – especially in the first half of the twentieth century – was an expensive affair, and an endeavor to record a trio could not have been easily expanded into a sextet, or big band. So, session degrees were preserved.

Second, I was preserving the number of sessions a musician played, with a window of one year. If a trumpet player recorded five times over the year when the session happened, I was allocating that player to five sessions in the rewired jazz world over that one-year period.[5]

---

[4] http://jazzprofiles.blogspot.com/2011/06/wynton-kelly-1931-1971-pure-spirit.html

[5] I was also generating random worlds with 2, 5 and 10 year windows, but the one year window is the closest to the logic of the jazz field, and it is also the strictest in the availability of musicians. Over five years a musician can completely change his or her style, and over ten years a musician might not even be alive.

Third, I was preserving the instrument combinations of sessions, as the recorded material would have been different without the same instruments.

Finally, I was only allocating musicians to a session to fill an instrument slot if they played the instrument over the current and previous year. Musicians often play multiple instruments, and it makes a considerable difference if a musician has played the instrument in question only a decade in the past.

In sum, I was generating jazz worlds where the recording company was able to fill the instrument slots in the session with available and able musicians. These musicians of course might have been very different from the observed ones. Imagine, for example Miles Davis on trumpet being swapped for a young trumpet player active in London at the time. The most important difference for us, of course, is that the musicians selected by our rewiring were not selected according to their networks. They were not likely to have played with each other in the past, and they were not likely to have been "friends of friends" either. I generated 100 jazz worlds – simulated complete histories of jazz where an observed jazz session had a corresponding rewired version.

**Prevalence of triad types in observed and rewired jazz worlds**

Figure 3 shows the relationship between edge weights and triadic closure, with areas corresponding to three kinds of triads, by showing the probability of triplet closure by minimal triplet leg weight quantiles. 5 338 093 Triplets were sliced into 10 000 quantiles of triplet legs weight. (There were over 5 million triplets in the observed data, and over 89 million triplets in the 100 random rewired jazz worlds combined.) Moving average smoothing was then applied to the curve of probability of closure. Quantiles range linearly on the x axis, from the first to the 10000$^{th}$. With quantiles charted on the horizontal axis and probabilities on the vertical axis, area on this chart is proportional to the number of triads.

The most apparent feature of Figure 3 is that the probability of closure is increasing with minimal triplet leg weight: if musician *i* played frequently in the past with musician *j*, and *j* played often with *k*, there is a higher probability that *j* and *k* have also played together at least once (compared to less frequent co-plays for *i-j* and *j-k* dyads). It is also apparent, that leg weights and the probability of closure in general is much lower in the rewired jazz worlds.

In the observed data (panel a of Figure 3) minimal triplets with legs weight that equals one are closed with .513 probability. The increase of closure probability is monotonic; the top percentile of minimal triplet legs weight (between 19 and 26) is closed with P=.994. In the rewired data, triplets with minimal legs weight has a probability of closure of .120, that increases suddenly around the threshold of forbidden triads of legs weight equals two (P=.418). The top percentile (minimal legs weight between 4 and 5) is closed with P=.778.

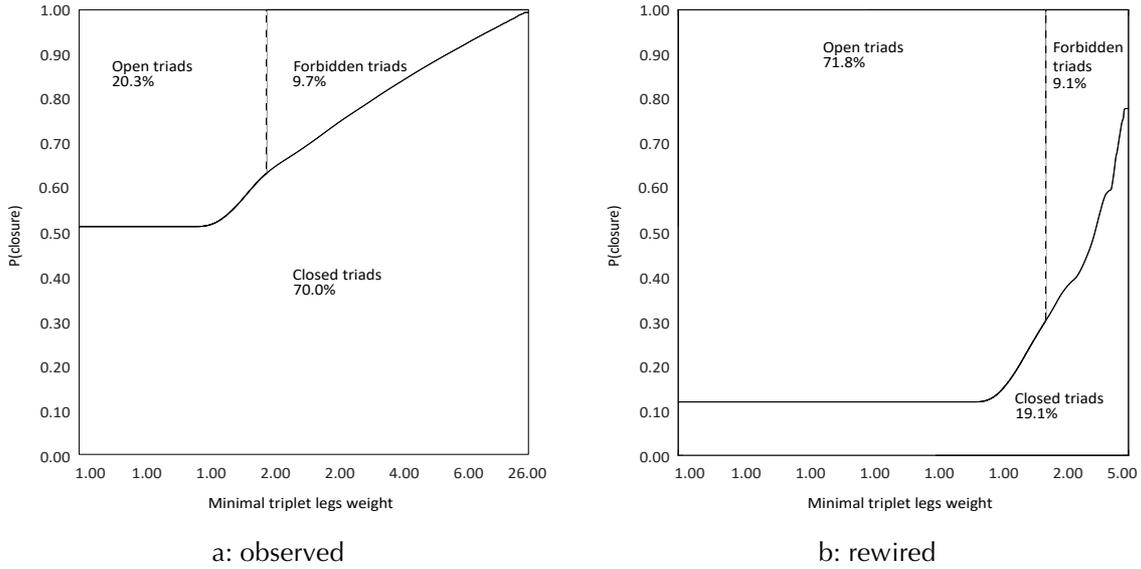

Figure 3: Area charts of the probability of triplet closure by minimal triplet legs weight quantiles.

Both in the observed and simulated data about 9% of triads are forbidden ones: triads with high edge weights without closure. In the observed data it seems that the number of forbidden triads is limited by the fact that the edge weights are relatively high, and high edge weight is correlated with closure as well. In the rewired data the number of forbidden triads seems to be limited by the fact that while most triads are open, the average edge weight is small.

To estimate the statistical significance of these differences, I computed a logistic regression model, where the dependent variable is the presence of closure ($y=1$), as opposed to an open triad ($y=0$), and the independent variables are: first, the triad belongs to the observed triads ($x_o = 1$), as opposed to the rewired (random jazz world) triads ($x_o = 0$), second, the minimal triplet legs weight ($x_w$), and third, the interaction between the observed triad indicator and minimal triplet legs weight ($x_o * x_w$). I was estimating the following logistic regression equation:

$$\ln \frac{P(y = 1|X)}{1 - P(y = 1|X)} = \alpha + \beta_o x_o + \beta_w x_w + \beta_{ow}(x_o x_w)$$

I estimated this logit model on a matched sample, where all of the observed triads are included (5,338,093 triads), and a sample of the same size is included from the rewired triads. There were 89,327,277 connected triads in the 100 random jazz worlds, I included a uniform probability random sample of 5,338,093 triads, to have 10,676,186 observations for the logistic regression estimation. The coefficient estimates are shown in Table 3. Since the units of observations are not independent (triads can share on or two edges), a permutation test was used to estimate p values for coefficients[6].

---

[6] A permutation test on the full sample was not practically feasible, so a random sample of 500 000 triads from both the observed, and the rewired triads was taken, and the model was estimated 10

|  | Beta | Odds ratio | SE | p |
|---|---:|---:|---:|---:|
| Observed | 2.712 | 15.054 | .003 | .000 |
| Min legs weight | 1.100 | 3.005 | .002 | .000 |
| Observed * Min legs weight | -.639 | .528 | .002 | .000 |
| Constant | -2.981 | .051 | .003 | .000 |
| N of observations | 10 676 186 | | | |
| Pseudo R-square | .302 | | | |
| Log likelihood | -5 122 486 | | | |

Table 3: Logistic regression estimation of triadic closure

The results show that the odds of closure in general is much higher (15.054 times higher) with observed triads than with rewired ones. In the real jazz world musicians are playing repeatedly with each other, while in the rewired world nothing ensures coherence – there are no band identities, and band leaders repeatedly playing with the same musicians. The estimated probability of closure for the weakest connected triad (with two edges of strength one, that is min leg weight of one) equals .549, while the same estimated probability for rewired triads is .133 (in line with the starting values on Figure 3). The odds of closure triples (multiplies by 3.005) with each additional increase in min leg weight for rewired triads, but for observed triads an increase in min leg weight only multiplies odds of closure by 1.587 (3.005*.528). Observed triads have a high baseline tendency to be closed, but they also have a tendency to "stay open longer" with increasing edge weights.

What this translates to as an actual process in jazz, is that players in the real jazz world can come together in a session where a musician had played a lot with one, and another fellow musician, yet these two alters have never seen each other in the studio. Forbidden triads seem to be a significant distinguishing feature of the jazz world, that does not show up in the randomly rewired version of this world. While the higher closure of the observed jazz world can easily be explained by factors washed out by the rewiring – such as clustering based on geographic proximity, style, and allegiance to a band leader – lower closure in high-weight triads is a non-trivial aspect of the observed jazz world.

The triads that we considered thus far were cut from their session context: the number and composition of musicians by their instruments, and the specific time of recording. The question that I answer subsequently is whether the proportion of forbidden triads at the level of sessions is different in the observed and rewired data. Thus far we reviewed the probability of closure by the weight of edges in triplet legs, without considering the session context. Our relevant unit of analysis for success is the session, and the question that I address is whether the prevalence of forbidden triads in observed sessions are different from rewired sessions. To answer this question, I compared the distribution of forbidden triads density in observed sessions to the average density that we see in 100 randomly rewired jazz worlds.

---

000 times with a randomly permuted dependent variable to generate a distribution of coefficients, and an estimate of p value for each.

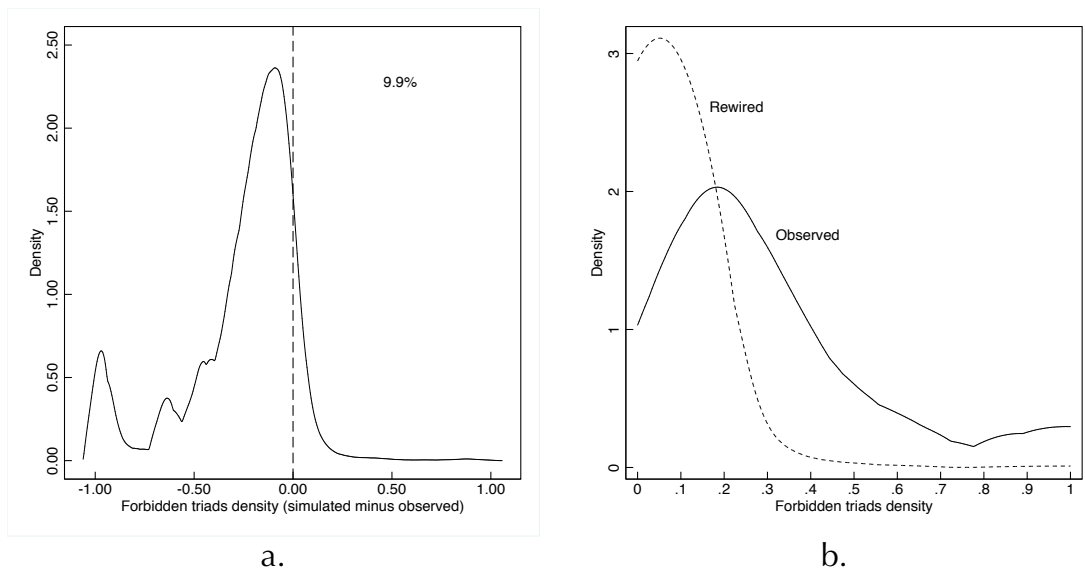

a.  b.

Figure 4: Kernel density estimation of: a.: rewired minus observed proportion of forbidden triads, and b.: the mean rewired proportion and the observed proportion of forbidden triads.

Figure 4 shows the distribution of the simulated minus observed proportion of forbidden triads, for sessions with at least one forbidden triad observed, and the overlaid density plots of rewired and observed densities of forbidden triads. Using kernel density estimation,[7] we see that most of the simulated sessions have a lower proportion of forbidden triads than the observed proportion. Only in 9.9% of sessions have we seen a proportion of forbidden triads that is higher than the observed. To test the null hypothesis that the two sets of forbidden triad proportions (simulated and observed) come from the same distribution, I used a Wilcoxon signed-rank test, and a Kolmogorov-Smirnov test for equality of probability distributions. Both of these tests fail to provide support for the null hypothesis of equal distributions. The Wilcoxon signed-rank test Z-score equals 124.7, with a corresponding p-value of 0.000. The Kolmogorov-Smirnov statistic equals 0.665, with a corresponding corrected p-value of 0.000. This supports the hypothesis that the proportion of forbidden triads in the observed jazz sessions are higher than what we would expect just based on the sizes of sessions, and the distribution of musician session participation.

**Regression models of success**

I used regression models to relate triad types to success. The dependent variable – the measure of success – in the regression models is the number of releases, ranging from one to 176. (The dependent variable in the OLS models is the ten-base logarithm of the number of releases ranging from .301 to 2.478.) Our key independent variables represent the density of three triad types – forbidden triads, closed triads, and open triads. Tests of multicollinearity are shown in Appendix A. Forbidden triads are defined as triads with a minimal triplet leg weight of two ($w_{(2)} \geq 2$), but the results presented are robust to varying this threshold. (See Appendix B for details on model robustness to varying $w_{(2)}$.) I am using the open triads category as the reference in the multivariate models, and enter squared terms for both closed and forbidden triads densities to capture non-linear (quadratic) relationships.

---

[7] I used an Epanechnikov kernel function, and an optimal bandwidth of 0.061 obtained by the normal scale rule (Härdle et al. 2004).

To mitigate biased estimates dues to right-censoring in the data, I excluded all sessions after the year 2000, leaving at least a ten-year window for each session to accumulate records. Unfortunately the release dates for releases is not available in the Tom Lord dataset, so I was unable to estimate the decay in the frequency of releases by time. I did however estimate models with varying time windows omitted from the right hand side (15 and 20 years), that did not affect the main findings. (See Appendix C for details on robustness to right hand side time cutoffs.)

Before specifying multivariate models, I turn to the bivariate relationships among triad densities and success. Figure 5 shows the bivariate quadratic relationship between three variables of triad density and the number of releases. All three triad types show an inverse U-shape relationship with the number of releases: the optimal density for each type of triad is around the middle of the range. To test the quadratic nature of these relationships I included categorical and lowess estimators of expected number of releases as well. For the categorical estimator I converted the interval scale triad density variables to four categories (0.00 to less than 0.25, 0.25 to less than 0.50, 0.50 to less than 0.75, and 0.75 to 1.00). The first inset on each panel shows the predicted number of releases by the four categories, with 95% confidence intervals. The second inset shows a locally weighted scatterplot smoothing estimation of the predicted number of releases, with a bandwidth $f=.5$, and with tricube weighting (Cleveland 1979). Vertical axis scales are not uniform in all insets to allow small figures to be more visible.

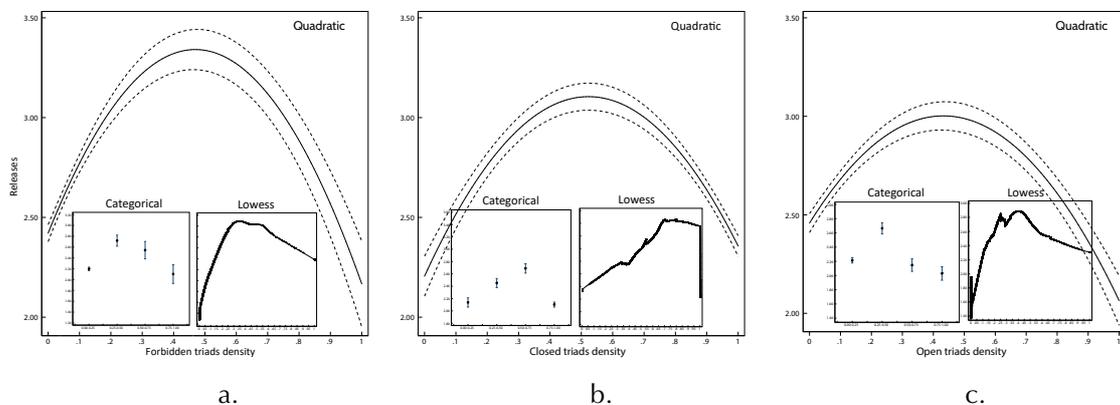

a.          b.          c.

Figure 5: Bivariate quadratic models of triad types and success, with 95% confidence intervals. Insets show categorical and lowess estimators.

All triad densities have an inverted U-shape relationship with success, suggesting that too little or too much of any kind of triad is a pathological social structure.

To test the appropriateness of the quadratic operationalization, I calculated the change in R-squared for models for each of the three triad shapes. For a given triad shape – say, the density of forbidden triads – I first estimated a model with a constant and the forbidden triad density variable. Then, a second model was estimated with the square of the forbidden triads density variables added, and I recorded the improvement of fit. I repeated this process up to the eighth power, and charted the resulting sequence of R-square improvements. I drew a line for each of the three triad shapes. The results are shown in Figure 6. For each triad the optimal model is the one with the quadratic term. Open triad models have an optimal fit with a cubic term, but the improvement over the quadratic term is small, and for the sake of model parsimony the first and second powers were included for all three triad shape variables.

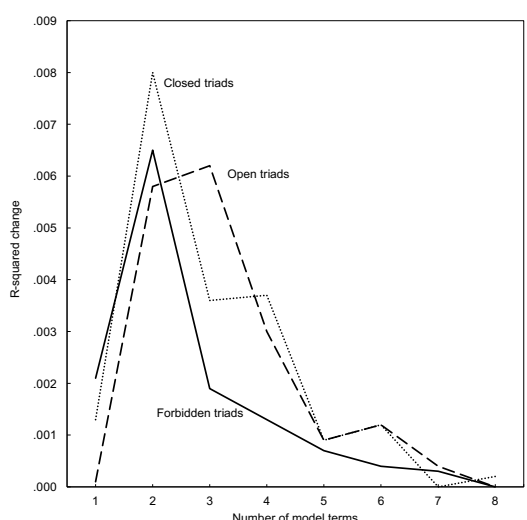

Figure 6: R-squared change in OLS predictions of the log number of releases by the number of model terms (powers) of triad density variables.

In the following analysis I am testing whether the advantage of forbidden triads over other triad shapes is statistically significant, and whether this relationship remains after controlling for other key alternative hypotheses. Further independent variables are entered, representing key alternative explanations to keep constant. A possible alternative explanation to forbidden triads can be the strength of ties in general. It is possible that forbidden triads are only significant predictors of success because they contain higher weight edges, and in fact it is the intensity of ties in the session that is related to success. Thus I entered the median tie strength in the session, and also the square of median tie strength to test for a non-linear relationship. It is reasonable to expect that at very high levels of tie strength musicians will not be as successful (the returns to tie strength is decreasing), as a high tie strength value might indicate getting locked into repeated and exclusive collaboration with the same set of musicians.

Instrumental innovation is another possible alternative explanation: the session is successful not because of a fold network dynamic, but because forbidden triads are a proxy for experimenting with a new instrument combination: Bringing in a new musician is really about bringing in a new instrument. Thus I entered distinctiveness: the average cosine distance of the instrument combination vector of the session (in the space of the top 200 most frequent instruments) to all other sessions over the preceding five years (t-1 to t-5).

The next four variables were entered to capture key attributes of the musicians in the session. The total number of musicians might be related to both success (a session employing more musicians can be more successful) and to the density of triad types (a large session can have a higher diversity of triad types). Newbies proportion is the proportion of musicians for whom the current session is the very first one. Employing beginner musicians might be correlated with lower closure (more open and forbidden triads), and might also be correlated with lower success, thus masking the relationship between triad types and success. The past success of musicians might be both a contributor of current success, and a reason for taking them on board despite a lack of prior ties with other musicians. And finally, the total amount of experience of musicians (measured in the total number of past sessions musicians played on) can contribute to both success and forbidden triads[8]. Table 4 shows Pearson correlation coefficients for all pairs of variables.

|  | 1. | 2. | 3. | 4. | 5. | 6. | 7. | 8. | 9. | 10. | 11. |
|---|---|---|---|---|---|---|---|---|---|---|---|
| 1. Releases | 1.000 | | | | | | | | | | |
| 2. log(Releases) | .822 | 1.000 | | | | | | | | | |
| 3. Forbidden triads | .026 | .046 | 1.000 | | | | | | | | |
| 4. Closed triads | -.018 | -.036 | -.510 | 1.000 | | | | | | | |
| 5. Open triads | .003 | .011 | -.066 | -.825 | 1.000 | | | | | | |
| 6. Median tie strength | .026 | .043 | -.036 | .165 | -.168 | 1.000 | | | | | |
| 7. Distinctiveness | -.159 | -.235 | -.070 | .106 | -.077 | -.017 | 1.000 | | | | |
| 8. Musicians (n) | .061 | .065 | .000 | -.072 | .084 | -.087 | -.097 | 1.000 | | | |
| 9. Newbies proportion | -.053 | -.086 | -.181 | .266 | -.189 | -.216 | .081 | .061 | 1.000 | | |
| 10. Median past releases | .683 | .663 | .024 | -.015 | .001 | .026 | -.187 | .082 | -.035 | 1.000 | |
| 11. Past sessions (n) | .025 | .065 | .252 | -.256 | .131 | .120 | -.160 | .224 | -.309 | .038 | 1.000 |
| 12. Year | -.360 | -.551 | -.005 | -.017 | .022 | -.026 | .306 | -.035 | -.037 | -.430 | .032 |

Table 4: Pearson correlation coefficients

I estimated both logged dependent variable OLS models, and negative binomial models appropriate for count dependent variables. It is most appropriate if statistical modeling of the number of releases takes into account the count nature of this dependent variable. Releases accumulate over time, and take on discrete positive values. Ordinary least squares models using the original releases variables are not entirely appropriate, as the distribution of count variables is more skewed than an assumed normal distribution. Negative binomial models are especially appropriate for over-dispersed count variables, which fits the number of releases accumulating in time (Cameron and Trivedi 1998; Greene 2003). Another possible approach is to use OLS regression on a logged dependent variable, that is less skewed than the original. To test model robustness I also used OLS models with a logged dependent variable. Using a negative binomial model has the advantage of taking overdispersion into account, and estimating a parameter of overdispersion explicitly, which is not available in OLS models of logged number of releases.

---

[8] Although tie strength and the number of past sessions are closely related, they do not measure the same thing. Imagine two sessions: one in which there is a high median tie strength, yet a relatively low number of past sessions, and one with zero median tie strength, and a high number of past sessions. In the first session musicians played all their past sessions with each other, while in the second one they accumulated diverse experiences.

I estimated these models with all session pooled, and with fixed effects included for band leaders, to account for leader specific unobserved heterogeneity in success. Beyond the independent variables presented above, both success and network structure in the session can be highly dependent on unobserved features of the band leader (or band identity in later decades). A session recorded by a high-reputation band leader can both attract high profile musicians with a proven track record and possibly a lack of closure (high forbidden triads density), and the same session might see more subsequent releases. Estimating models that keep the identity of the leader constant (estimate within-leader effects) can get us closer to the goal of capturing the impact of network structure on success. I followed the derivation of the fixed effects negative binomial model proposed by Hausman and coauthors (Hausman, Hall, and Griliches 1984), and also include fixed effects OLS models for logged number of releases. This is especially useful in light of more recent doubts about the Hausman-model, namely that the derivation builds the fixed-effects into the distribution of the gamma heterogeneity, α, not the mean (Allison and Waterman 2002).

Starting with the simplest of the estimators, an OLS models with a logged dependent variable estimates the following equation:

$$\log y_s = \beta' X_s + \varepsilon_s$$

for $s=1,2,…N_{sessions}$, where $y_s$ is the number of releases that resulted from the session, $\beta$ are model parameters, $X_s$ are the set of independent variables, and $\varepsilon_s$ is the error term, and estimation is ordinary least squares. The same model with band leader fixed effects would be written as:

$$\log y_s = \gamma_l + \beta' X_{ls} + \varepsilon_{ls}$$

for $l=1,2,…N_{leaders}$, and $s=1,2,…N_{sessions}$, where an additional term, $\gamma_l$ estimates the band leader specific fixed effect (a band leader specific baseline level of success).

The negative binomial model estimates the occurrence rate of releases, and takes into account the nature of count data as non-negative integers. The negative binomial model without fixed effects is estimated by the following equation:

$$\ln \mu_s = \beta' X_s + \varepsilon_s$$

for $l=1,2,…N_{leaders}$, where $\mu_s = E(y_s|x_s)$ is the occurrence rate of releases, $\beta$ are model parameters, $X_s$ are the set of independent variables, $\varepsilon_s$ is the error term, and the model is based on a maximum likelihood estimator. The negative binomial model is an extension, or rather a generalized version of Poisson regression models, in that it estimates the over-dispersion parameter $\alpha$, that equals zero for the special case that is a Poisson model (with no overdispersion, a mean that equals variance). The null hypothesis that $\alpha$ equals zero is testable.[9]

The equation for a negative binomial model with fixed effects is then accordingly:

$$\ln \mu_{ls} = \gamma_l + \beta' X_{ls} + \varepsilon_{ls}$$

for $l=1,2,\ldots N_{\text{leaders}}$, and $s=1,2,\ldots N_{\text{sessions}}$, where $\mu_{ls} = E(y_{ls}|x_{ls})$ is the occurrence rate, $\gamma_l$ is the band leader specific fixed effect, $\beta$ are model parameters, $X_{ls}$ are the set of independent variables, and $\varepsilon_{ls}$ is the error term.

**Predictors of success**

Table 5 summarizes the results of these statistical models. The first two models are estimated without fixed effects for band leaders, while the third and fourth models include fixed effects for band leaders. The first and third models are OLS models of log number of releases, the second and fourth models are negative binomial (NB) models of the number of releases. The test of overdispersion in the negative binomial models indicate that the occurrence rate of releases is significantly more dispersed than a Poisson distribution ($\alpha =.257$, with p<.000), warranting the negative binomial specification.

The coefficient of the first power of forbidden triads density is positive and significant, while the second power is negative and significant. This suggests that – compared to open triads – an increase in forbidden triads can initially be more beneficial for the number of releases. The density of closed triads is not different from open triads with regards to releases. These findings are consistent across all four models, with or without band leader fixed effects, and with OLS and NB specifications.

This suggests that comparing two sessions of the same size (the same number of triads), the one that features more forbidden triads (at least to a limit of forbidden triads proportion) can be more successful than the session that has only closed and open triads. The paradoxical triad of strong-strong-absent ties seems to be the best predictor of success – even if we compared sessions of the same band leader.

Of the control variables only three are consistent across all four models: Having a higher proportion of newbies means a lower number of releases. A better past track record (higher median number of releases for the past recordings of the musicians on the session) means a higher number of releases to the target session. The overall number of releases decreases as years advance – indicating either a long-range censoring, or a general decline in the level of success for the average jazz session over time.

---

[9] See (Cameron and Trivedi 1998) for formulas and estimation of the $\alpha$ parameter.

|                                | 1. OLS model of log(Releases) | 2. NB model of Releases | 3. OLS model of log(Releases) with fixed effects | 4. NB model of Releases with fixed effects |
|---|---|---|---|---|
| Forbidden triads | .0445*** (.0100) | .2648*** (.0511) | .0316*** (.0112) | .2253*** (.0528) |
| Forbidden triads (squared) | -.0383*** (.0114) | -.2282*** (.0585) | -.0291** (.0127) | -.2672*** (.0606) |
| Closed triads | -.0052 (.0086) | -.0252 (.0449) | .0147 (.0099) | .0730 (.0471) |
| Closed triads (squared) | .0044 (.0078) | .0114 (.0409) | -.0173 (.0090) | -.0843* (.0429) |
| Median tie strength | -.0007*** (.0002) | .0071*** (.0012) | -.0006* (.0003) | -.0071*** (.0013) |
| Median tie strength (squared) | .0000* (.0000) | -.0001*** (.0000) | -.0000 (.0000) | -.0000 (.0000) |
| Distinctiveness | -.0723*** (.0067) | -.5120*** (.0354) | -.0425*** (.0094) | .1248*** (.0428) |
| Musicians (n) | .0002*** (.0001) | -.0024*** (.0005) | .0011*** (.0001) | .0078*** (.0006) |
| Newbies proportion | -.0779*** (.0033) | -.5364*** (.0185) | -.0783*** (.0047) | -.2066*** (.0224) |
| Median past releases | .0358*** (.0002) | .1162*** (.0010) | .0261*** (.0002) | .0239*** (.0003) |
| Past sessions (n) | -.0000*** (.0000) | .0001*** (.0000) | -.0000*** (.0000) | -.0005*** (.0000) |
| Year | -.0040*** (.0000) | -.0200*** (.0001) | -.0055*** (.0001) | -.0027*** (.0004) |
| Constant | .7215*** (.0049) | 2.3051*** (.0261) | -.8403*** (.0088) | 1.5289*** (.0392) |
| Fixed effects for band leader | No | No | Yes | Yes |
| N of observations | 81527 | 81527 | 72042 | 72042 |
| F | 7866.36*** |  | 1886.44*** |  |
| Chi-square |  | 65153.39*** |  | 8088.77*** |
| R-square (adjusted) | .536 | .189[1] | .514 | .339[1] |
| Log likelihood |  | -139558.13 |  | -104666.63 |

*Notes*: 1: McFadden's adjusted pseudo R-squared is used. Standard errors are in parentheses. *: P<.05; **: p<.01; ***: p<.001.

Table 5: Statistical models of success.

To better grasp the quadratic relationship between triad types and success I charted the marginal effects of forbidden triad density and closed triad density on the number of releases for all four models. Marginal effects plots show the predicted levels of the dependent variable as we vary one given independent variable – a triad density – and keep all other variables fixed at their mean values. These charts allow us to manipulate an imaginary session where every aspect – including the identity of the band leader for fixed effects specifications – are the same, but the triad density in question is changing. Figure 7 displays two marginal effects plots for each of the four models.

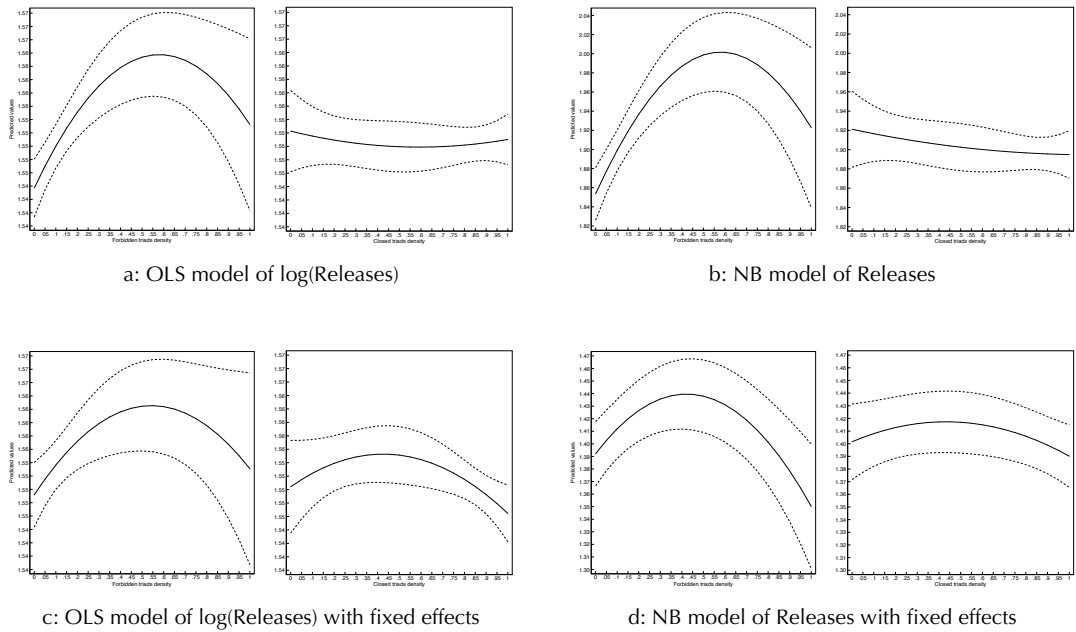

Figure 7: Marginal prediction of number of releases for triad shapes, with 95% confidence intervals.

All four models show the same pattern for both the density of forbidden triads, and the density of closed triads. Increasing forbidden triads density increases the predicted number of releases as well, with a peak of around .50 forbidden triad density (half of the connected triads are forbidden triads). The particular initial and peak predicted number of releases vary from model to model, and the predicted boost to the number of releases just from moving from no forbidden triads to the optimal range is between one and eight percent. There is no evidence, however, to any benefits from closure. All four models show basically a flat line for closed triad density (or a line well within the initial confidence intervals). The reference category is the density of open triads (the three densities sum to one); these findings thus show that forbidden triads outperform both open and closed triads in their mid-region.

Figure 8 shows similar plots of margins for the Median tie strength variable. As opposed to the results about forbidden triads, there is no consistent relationship between Median tie strength and the number of releases.

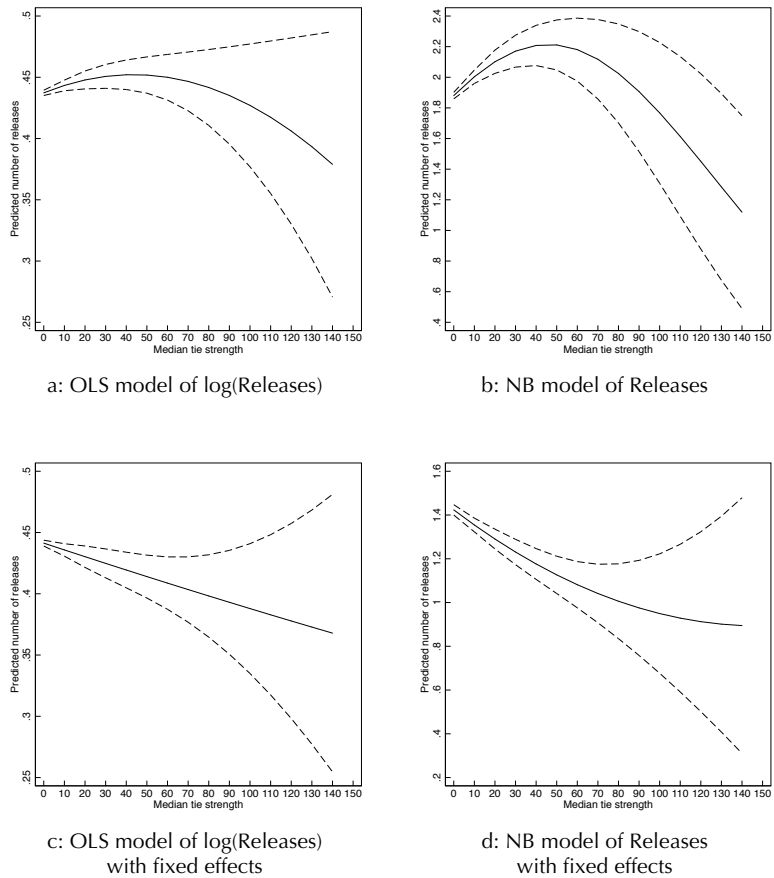

Figure 8: Marginal prediction of number of releases for median tie strength, with 95% confidence intervals.

Returning to the example of Miles Davis, I ran the negative binomial version of our regression model with a constant, linear, and quadratic interactions included for his sessions. This enables us to chart the way in which forbidden triads predicted success for the most iconic of all of jazz's band leaders. Figure 9 shows the difference between Miles's sessions and all other ones. Forbidden triads seem to have contributed more to the success of Miles Davis's sessions than to all other sessions, on average. While the higher baseline success might not be surprising (this was the motivation for our fixed effects operationalizations), the faster accelerating curve for Miles is clearly different from the rest. It is not only that a Davis session is more successful on average, but his sessions got more success out of forbidden triads than others. (Note that the figure only displays forbidden triads up to .50, as the standard errors drastically increase for the sessions of Miles Davis after that point: there were simply too few sessions to make meaningful predictions about the higher range.)

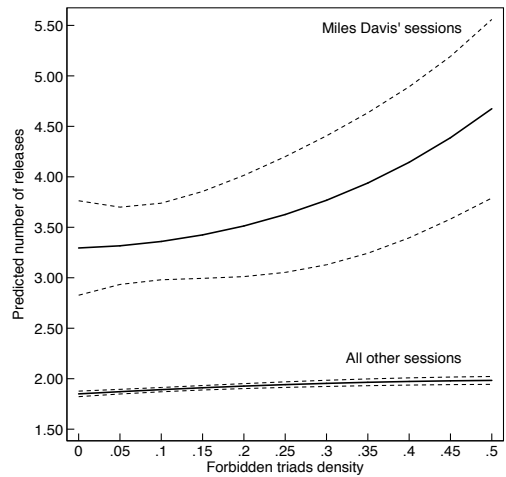

Figure 9: Marginal effects plot for forbidden triads density in Miles Davis' sessions using the NB model.

The example of Miles Davis indicates that high-profile band leaders might take special advantage of network structures, such as forbidden triads. This raises the question whether the findings of our statistical modeling is robust to omitting high profile band leaders. To test this, I omitted 131 most outstanding jazz artists – those that received the National Endowment of the Arts Jazz Masters nomination.[10] The results without the sessions of the outstanding band leaders are practically identical to the results in the full dataset. (See Appendix D for details.)

To summarize, jazz sessions are more successful if musicians have forbidden triads in their collaboration network: if there is a diversity of socially evolved styles, striking a productive balance between familiarity and freshness. It is important to contrast the consistent finding about the significance of forbidden triads to the ambiguous evidence for the importance of instrument combinations. Distinctiveness in combining instrument is a negative predictor of success in three out of the four models, contrary to the perceived image of jazz as a domain of constant experimentation. It seems that an unexpected mix of instruments is not enough to generate a new sound. Forbidden triads are about a more subtle, more social kind of experimentation that can tap into a novel combination of socially evolved styles rather than merely a combination of instruments.

**Conclusions**

The history of the jazz field is about constant experimentation – a quest for new sounds. This paper is about the source of new sounds that make a recording successful. I tested hypotheses about the structure of the collaborative network, and hypotheses about attributes of musicians, and the session. The most promising alternative explanation about the importance of instrument combinations was not supported by our regression models. A distinctive instrumentation is a liability, if anything: most models show a significant negative coefficient. Our analysis suggests a different source for new sounds, based on the combination of jointly evolved styles, rather than simply instruments.

---
[10] https://www.arts.gov/honors/jazz

I represented a sessions' collaboration network by the density of triad types. The general imagery of social networks parses triads into two kinds: closed triads that make up cohesive clusters, and open triads that bridge between these clusters. Strong ties are seen to be related with closure, and weak ties are seen as the building blocks of open triads. The third possibility: the openness of strong ties is seen as anomalous, a rare and residual category, a triad that is forbidden (Granovetter 1973). Thus far no one questioned the intuition of Granovetter from more than four decades ago: forbidden triads have not been investigated as network structures of interest.

I argue that forbidden triads are of crucial importance to understand innovation. Novelty is about an interplay of familiar and novel, about the interplay of trusted ties and an unfamiliar face. A forbidden triad is a molecule of an innovative network. In that triad two familiar dyads meet at one of their nodes. A musician invites two co-players to play together for the first time. Strangers are also meeting for the first time, but they have no familiarity to start from. The central node in a forbidden triad can mobilize the trust that he or she has with both of the alters, and he or she can start the translating work between the two jointly evolved styles that can lead to a new sound.

**Appendix A: Multicollinearity**

While multicollinearity can be of concern in general for multivariate statistical models, understanding the robustness of the findings about forbidden triads to multicollinearity is especially important here, as the proportion of forbidden triads and the proportion of closed triads can not vary completely independently, by definition. To assess the general level of multicollinearity for all independent variables, I include table 5, with variance inflation factors. The VIF values indicate that multicollinearity is not a concern – all statistics are well within customary bounds (of VIF>=4.00).

|  | VIF |
| --- | --- |
| Forbidden triads | 1.39 |
| Closed triads | 1.52 |
| Median tie strength | 1.13 |
| Distinctiveness | 1.15 |
| Musicians (n) | 1.10 |
| Newbies proportion | 1.25 |
| Median past releases | 1.24 |
| Past sessions (n) | 1.29 |
| Year | 1.33 |

Table 6: Variance inflation factors for independent variables.

To test sensitivity to the specific dependence among the proportion of forbidden triads and closed triads (the maximum of one cannot be greater than one minus the value of the other), I re-computed all models omitting the proportion of closed triads variable. These models show the same prediction for forbidden triads. The full models are omitted, but available from the author upon request. The resulting predicted margins for the proportion of forbidden triads are shown on Figure 10. The conclusions are the same as for the models in Table 5, and the results are consistent throughout the four modeling approaches.

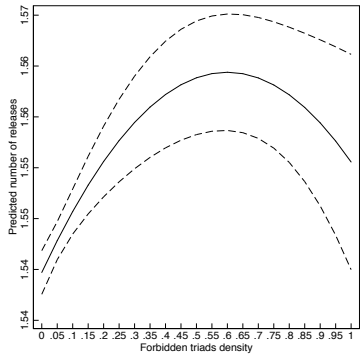
a: OLS model of log(Releases)

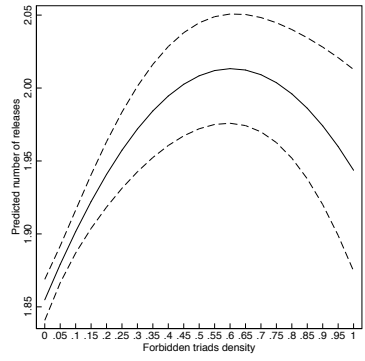
b: NB model of Releases

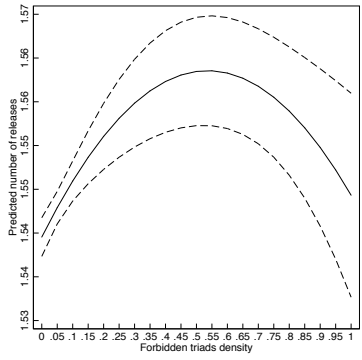
c: OLS model of log(Releases)
with fixed effects

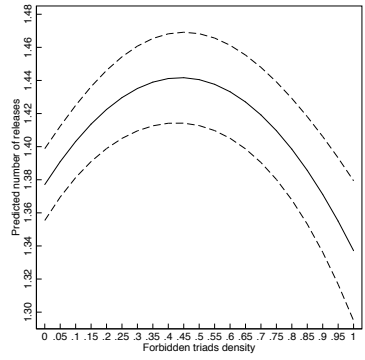
d: NB model of Releases
with fixed effects

Figure 10: Marginal prediction of number of releases for forbidden triads when closed triads are omitted, with 95% confidence intervals.

## Appendix B: Robustness to minimal triplet legs weight threshold

Since the definition of forbidden triad depends on the threshold of minimal triplet leg weight, I tested the sensitivity of our findings to varying this threshold. The original definition relies on the minimal threshold of $w_{(2)} \geq 2$, since tie weights are integer (recording the number of prior co-plays). In this appendix I test thresholds of 3, 5, and 10 as well. Figure 11 shows the bivariate quadratic relationship between forbidden triads density and number of releases. The shape of the curves with the 3, 5 and 10 thresholds are very similar.

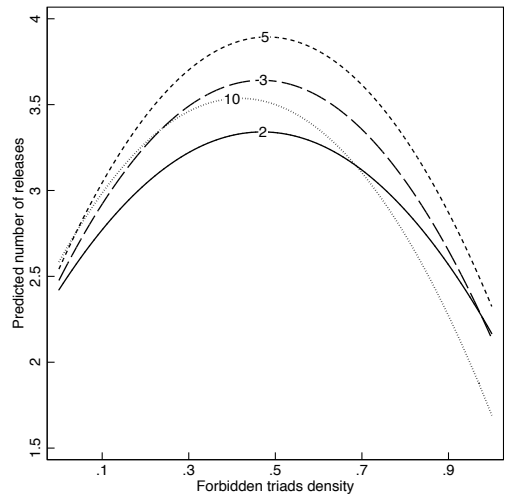

Figure 11: Bivariate quadratic relationship between forbidden triads density and number of releases, with various thresholds for minimal triplet legs weight. Lines are labeled by the threshold value.

I ran all the models shown in Table 5 with the 3, 5, and 10 thresholds as well. Table 7 shows only the coefficients for forbidden triads density, and squared forbidden triads density. The models were the same as the ones shown in Table 5; all other variables are omitted to save space. The original threshold of $w_{(2)} \geq 2$ is shown for comparison at the top of Table 7.

| Minimal triplet legs weight threshold | | 1. OLS model of log(Releases) | 2. NB model of Releases | 3. OLS model of log(Releases) with fixed effects | 4. NB model of Releases with fixed effects |
|---|---|---|---|---|---|
| $w_{(2)} \geq 2$ | Forbidden triads | .0445*** (.0100) | .2648*** (.0511) | .0316*** (.0112) | .2253*** (.0528) |
| | Forbidden triads (squared) | -.0383*** (.0114) | -.2282*** (.0585) | -.0291** (.0127) | -.2672*** (.0606) |
| $w_{(2)} \geq 3$ | Forbidden triads | .0413*** (.0124) | .3099*** (.0617) | .0322* (.0136) | .2145*** (.0641) |
| | Forbidden triads (squared) | -.0368** (.0146) | -.2643*** (.0739) | -.0321* (.0160) | -.2842*** (.0766) |
| $w_{(2)} \geq 5$ | Forbidden triads | .0444* (.0197) | .2798** (.0966) | .0420* (.0112) | .1976* (.0996) |
| | Forbidden triads (squared) | -.0366 (.0242) | -.1885 (.1193) | -.0422 (.0259) | -.2817* (.1238) |
| $w_{(2)} \geq 10$ | Forbidden triads | .0247 (.0528) | -.1103 (.2605) | .0574 (.0554) | .3009 (.2639) |
| | Forbidden triads (squared) | -.0322 (.0628) | .0308 (.3211) | -.0892 (.0652) | -.5711 (.3188) |

Standard errors are in parentheses. *: P<.05; **: p<.01; ***: p<.001.

Table 7: Forbidden triads coefficients from statistical models of success with varying minimal triplet legs weight thresholds.

The threshold of 3 shows almost identical results to the original threshold of 2. With the threshold set at 5, the quadratic term is not significant (except in the fixed effects negative binomial model), while the first term is significant and very similar to prior models. This suggests that as the definition of forbidden triad becomes more restrictive (and the number of sessions with a high density of forbidden triads become very small), only the first part of the curve is there: One can identify higher success in sessions with moderate levels of forbidden triads density (compared to no forbidden triads), but the lower level of success in sessions with extreme densities of forbidden triads is no longer statistically significant. At the threshold of $w_{(2)} \geq 10$ the standard errors increase further, and the relationship between forbidden triad density and success becomes insignificant.

## Appendix C: Robustness to varying right-hand time cutoffs

The dependent variable – the number of releases subsequent to the recording of the session material – is sensitive to right-hand censoring, as releases accumulate with the passing of time. A recent session can be very successful, but we just had no chance to observe this success in a high number of releases yet. The regression models in Table 5 omit data within a ten year window on the right: sessions between 2000 and 2010 are omitted. To assess the sensitivity of the regression models to varying this time window, I include versions of these models with a fifteen and a twenty year window as well, omitting data after 1995, and 1990 respectively. Table 8 shows the coefficients for the first and second power of forbidden triads density only; other variables and model statistics are omitted to save space.

| Right-hand time cutoff | | 1. OLS model of log(Releases) | 2. NB model of Releases | 3. OLS model of log(Releases) with fixed effects | 4. NB model of Releases with fixed effects |
|---|---|---|---|---|---|
| 2000 | Forbidden triads | .0445*** (.0100) | .2648*** (.0511) | .0316*** (.0112) | .2253*** (.0528) |
| | Forbidden triads (squared) | -.0383*** (.0114) | -.2282*** (.0585) | -.0291** (.0127) | -.2672*** (.0606) |
| 1995 | Forbidden triads | .0476*** (.0118) | .2674*** (.0553) | .0342** (.0130) | .3437*** (.0406) |
| | Forbidden triads (squared) | -.0401** (.0135) | -.2262*** (.0636) | -.0320* (.0149) | -.3399*** (.0471) |
| 1990 | Forbidden triads | .0530*** (.0142) | .2771*** (.0606) | .0385* (.0155) | .3581*** (.0419) |
| | Forbidden triads (squared) | -.0425** (.0163) | -.2261** (.0699) | -.0347* (.0178) | -.3507*** (.0487) |

Standard errors are in parentheses. *: P<.05; **: p<.01; ***: p<.001.

Table 8: Forbidden triads coefficients from statistical models of success with varying right-hand time cutoff.

Our findings are not influenced by censoring on the right hand side: All models with varying cutoffs (2000, 1995, 1990) show the same results, with the same coefficients being significant. It is only the baseline level of success that is increasing as one is increasing the cutoff, as older sessions had more time to accumulate releases. Figure 12 shows the margins for the fixed effects OLS model (model 3), with all variables fixed at their means.

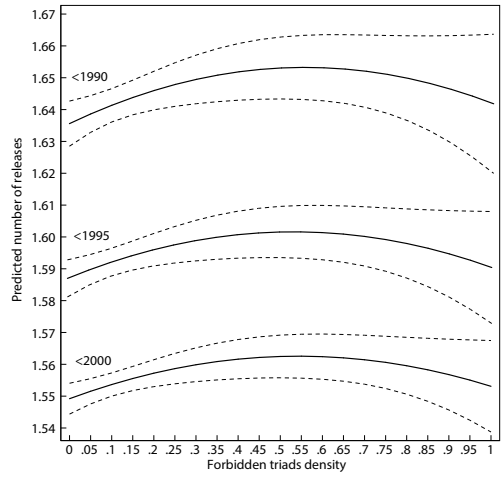

Figure 12: Marginal effects for forbidden triads density using the fixed effects OLS specification for three right-hand time cutoffs

**Appendix D: Robustness to omitting high-reputation leaders**

Success in jazz does depend greatly on the reputation of the band leader, which might greatly influence the finding that forbidden triads density is related to the number of releases stemming from the session. The results of the regression models might be greatly influenced by the fact that high reputation band leaders are more likely to attract active musicians who have not played with each other before, because the band leader's reputation goes beyond a connected network neighborhood. At the same time high reputation band leaders might also be more likely to have their material included on a higher number of releases. Fixed effects models do take band leader specific unobserved variance into account to some extent: these models add leader-specific constants.

As an additional test, I ran our models without band leaders with the highest reputations. To capture high reputation, I used the set of leaders that received the National Endowment of the Arts Jazz Masters nomination[11]: "the highest honor that our nation bestows on jazz artists," with recipients such as Miles Davis, Ella Fitzgerald, Herbie Hancock, Wayne Shorter, and Dave Brubeck. There were altogether 145 artists awarded, and I found 131 of them as band leaders in our dataset (the 14 remaining artists had not been band leaders, or our datasets did not contain any sessions from them). I excluded all the 7,963 sessions of the 131 awarded leaders from our dataset, and re-ran our models. Table 8 shows the resulting models, presented in an identical way to the models in Table 5. The models stayed the same: coefficients of forbidden triads density are almost identical, the only difference being that in model 3 the squared term is not significant: in the fixed effects OLS model there is no evidence for a detrimental effect of very high forbidden triads density. We can safely conclude that the evidence for the relationship between forbidden triads and success is independent of the presence of high reputation band leaders.

---

[11] https://www.arts.gov/honors/jazz

|  | 1. OLS model of log(Releases) | 2. NB model of Releases | 3. OLS model of log(Releases) with fixed effects | 4. NB model of Releases with fixed effects |
| --- | --- | --- | --- | --- |
| Forbidden triads | .0372*** | .2503*** | .0256* | .3124*** |
|  | (.0099) | (.0533) | (.0112) | (.0432) |
| Forbidden triads (squared) | -.03170** | -.2334*** | -.0226 | -.2905*** |
|  | (.0113) | (.0612) | (.0127) | (.0499) |
| Closed triads | -.0068 | -.0364 | .0127 | .1828*** |
|  | (.0084) | (.0464) | (.0099) | (.0388) |
| Closed triads (squared) | .0058 | .0150 | -.0152 | -.1948*** |
|  | (.0077) | (.0423) | (.0090) | (.0352) |
| Median tie strength | .0006* | .0070*** | -.0005 | -.0001 |
|  | (.0002) | (.0013) | (.0003) | (.0010) |
| Median tie strength (squared) | .0000 | -.0001** | -.0000 | -.0000 |
|  | (.0000) | (.0000) | (.0000) | (.0000) |
| Distinctiveness | -.0606*** | -.4545*** | -.0347*** | .0989** |
|  | (.0065) | (.0364) | (.0094) | (.0369) |
| Musicians (n) | .0002** | -.0025*** | .0011*** | .0078*** |
|  | (.0001) | (.0005) | (.0001) | (.0006) |
| Newbies proportion | -.0691*** | -.4854*** | -.0703*** | -.4646*** |
|  | (.0032) | (.0188) | (.0046) | (.0186) |
| Median past releases | .0399*** | .1238*** | .0291*** | .0313*** |
|  | (.0002) | (.0011) | (.0002) | (.0002) |
| Past sessions (n) | .0000*** | .0001*** | -.0000*** | -.0001*** |
|  | (.0000) | (.0000) | (.0000) | (.0000) |
| Year | -.0037*** | -.0188*** | -.0049*** | -.0325*** |
|  | (.0000) | (.0002) | (.0001) | (.0004) |
| Constant | .6713*** | 2.1477*** | .7735*** | 25.7289*** |
|  | (.0049) | (.0270) | (.0091) | (5.8185) |
| Fixed effects for band leader | No | No | Yes | Yes |
| N of observations | 75968 | 75968 | 66486 | 66486 |
| F | 7866.36*** |  | 1886.44*** |  |
| Chi-square |  | 59960.04*** |  | 26950.60*** |
| R-square (adjusted) | .554 | .192[1] | .534 | .342[1] |
| Log likelihood |  | -126420.48 |  | -100230.29 |

*Notes*: 1: McFadden's adjusted pseudo R-squared is used. Standard errors are in parentheses. *: p<.05; **: p<.01; ***: p<.001.

Table 8: Statistical models of success, with National Endowment for the Arts Jazz Masters award winning band leaders excluded.